\documentclass[useAMS,usenatbib,usegraphicx]{mn2e}
\newcommand{\lya}{Ly$\alpha$}
\newcommand{\cmmt}{{\rmn{cm^{-2}}}}
\newcommand{\mpc}{{\rmn{Mpc}}}

\newcommand{\kms}{{\rmn{km~s^{-1}}}}
\newcommand{\ang}{{\rmn{\AA}}}
\newcommand{\kelvin}{{\rmn{K}}}

\newcommand{\hi}{{\mbox{H\,{\sc i}}}}
\newcommand{\hii}{{\mbox{H\,{\sc ii}}}}
\newcommand{\hei}{{\mbox{He\,{\sc i}}}}
\newcommand{\heii}{{\mbox{He\,{\sc ii}}}}
\newcommand{\heiii}{{\mbox{He\,{\sc iii}}}}
\newcommand{\popii}{{\mbox{Pop\,{\sc ii}}}}
\newcommand{\popiii}{{\mbox{Pop\,{\sc iii}}}}
\newcommand{\dd}{{\rmn{d}}}
\newcommand{\fnu}{{\rmn{erg~s^{-1}~cm^{-2}~Hz^{-1}}}}

\newcommand{\nhei}{N_\rmn{HeI}}
\newcommand{\nhi}{N_\rmn{HI}}
\newcommand{\etathin}{\eta_\rmn{thin}}

\title[The UV background after reionization]{A method to infer the
stellar population that dominated the UV background at the end of
reionization}

\author[M. R. Santos and A. Loeb]{Michael R. Santos$^{1}$
\thanks{E-mail: mrs@tapir.caltech.edu (MRS); loeb@cfa.harvard.edu (AL)} 
and Abraham Loeb$^{2,3}$\\
$^{1}$Theoretical Astrophysics, 130-33 Caltech, Pasadena, CA 91125, USA\\
$^{2}$Astronomy Department, Harvard University, 60 Garden St.,
Cambridge, MA 02138, USA\\
$^{3}$Guggenheim fellow; currently on sabbatical leave at the Institute for
Advanced Study, Princeton, NJ 08540, USA}

\begin{document}

\date{\today}

\pagerange{\pageref{firstpage}--\pageref{lastpage}} \pubyear{2002}

\maketitle

\label{firstpage}

\begin{abstract}
We present an observational test of the spectrum of the ionizing
background at $z \simeq 5$; the test is sufficiently sensitive to
determine whether \popii\ or \popiii\ stars are the dominant source of
ionizing radiation.  The ionizing background at $z\simeq5$ may reflect
the nature of the sources responsible for the final overlap phase of
reionization.  We find that rest-frame extreme-UV \hei\ absorption
will be detectable in deep spectral observations of some rare
$z\simeq5$ quasars; the ratio of \hei\ to \hi\ absorption reflects the
shape of the ionizing background in the photon energy range between
13.6 and 24.6~eV.  Most $z\simeq5$ quasars have too much \hi\
absorption along their line of sight for \hei\ absorption to be
observed.  However, based on current measurements of \hi\ absorber
statistics, we use Monte Carlo simulations to demonstrate that the
Sloan Digital Sky Survey (SDSS) will discover a sufficient number of
$z \simeq 5$ quasars to turn up a quasar suitable for measuring \hei\
absorption (and we illustrate a selection method to identify that
quasar).  From simulated observations of a suitable $z\simeq5$ quasar
with a 10-meter telescope, we show that a constraint on the spectral
slope of the ionizing background at that redshift can be obtained.
\end{abstract}

\begin{keywords}
quasars: absorption lines -- intergalactic medium -- diffuse radiation
-- cosmology: theory
\end{keywords}

\section{Introduction}
\label{sec:intro}

Intergalactic hydrogen is almost completely ionized by $z\simeq 6$
\citep{bec01,djo01}, but the sources responsible for reionization are
still not known (see \citealt{bar01} and \citealt{loe01} for reviews
of the theoretical possibilities).  A low neutral-hydrogen fraction is
found at lower redshift even in relatively over-dense systems, despite
the short (compared to the Hubble time) recombination time-scale in
those systems.  This implies a continued supply of ionizing photons
after reionization completed.  Measurements of the intensity of the
ionizing background below $z \sim 3$ \citep{sco00} are consistent with
an ionizing background produced by observed quasars \citep{haa96}.  At
higher redshifts, however, the declining abundance of bright quasars
\citep[e.g.,][]{fan01b} suggests that they cannot provide an ionizing
background sufficient to reionize the universe \citep{wyi03a}, and
this conclusion is supported by deep x-ray surveys \citep{barg03}.

Recent results from the \textit{WMAP} satellite imply that
reionization was substantially underway by $z\sim15$
\citep{kog03,spe03}.  However, quasar absorption studies at $z\simeq6$
\citep{bec01,djo01} suggest a dramatic evolution in the ionizing
background around that redshift.  Some authors have reconciled the
\textit{WMAP} evidence for early reionization with the quasar result
that reionization is just finishing at $z\simeq6$ by postulating two
reionizations of the universe \citep{cen03a,cen03b,wyi03a,wyi03b},
generally invoking a hard \popiii\ stellar spectrum for the early
reionization and a \popii\ stellar spectrum for the final reionization
at $z\simeq6$.

The most direct technique to ascertain the sources responsible for the
ionizing background at high redshift is to take a census of all
sources that contribute ionizing photons.  The advantage of this
technique is that the relative contribution of different types of
sources are measured directly, as are the spatial distributions and
other properties of the populations.  Unfortunately, the large
luminosity distance to high redshift means that only the most luminous
sources can be detected with current observational methods.  For
example, \citet{fan01b} measure a power-law high-redshift quasar
luminosity function with a steep slope, but estimate that the
traditional power-law break of the quasar luminosity function would
occur more than one magnitude below their survey limit: there is
little constraint on the faint end of the quasar luminosity function
from direct source detection.

A complementary method to direct detection is measurement of the
properties of the integrated ionizing background.  Although this
technique does not identify ionizing sources directly, it is effective
at detecting ionization due to a very abundant population of
low-luminosity sources.  Moreover, the shape of the integrated
ionizing background is dictated by the combination of sources
responsible for it; different spectra arise from various possible
contributors, such as quasars \citep{tel02}, \popii\ OB stars
\citep[e.g.,][]{lei99}, very massive metal-free (\popiii) stars
\citep*{bro01}, and x-rays from early stars \citep{oh01}.

This paper describes possible observations that would constrain the
spectral shape of the ionizing background just after reionization
finished.  In Section \ref{sec:absspec}, we introduce quasar
absorption line spectroscopy in the context of this paper.  Section
\ref{sec:ionback} describes our parametrization of the ionizing
background, and the resulting ionization state of absorption systems
in the intergalactic medium (IGM).  In Section \ref{sec:igmabs} we
apply those results to observed statistics of \hi\ absorbers to
generate a model for \hi\ and \hei\ opacity toward high-redshift
quasars; the assumed properties of the quasars are given in Section
\ref{sec:qsospec}.  The method of simulating line of sight (LOS)
absorption spectra and studying the sensitivity of the results to the
shape of the ionizing-background spectrum is described in Section
\ref{sec:losqso}.  Section \ref{sec:qsosel} presents a method to
select the best quasars for absorption line spectroscopy.  We
summarize our results in Section \ref{sec:summ}.

\section{Absorption spectroscopy}
\label{sec:absspec}

The easiest way to make a crude measurement of the shape of the
ionizing background is through the ionization state of photoionized
IGM gas; in the next Section we show explicitly how the background
spectrum is related to the ionization state.  Atoms and ions with at
least one bound electron in the ground state have strong bound-bound
transition cross-sections, typically located at UV or soft x-ray
energies.  Consequently, even small amounts of these species on the
line of sight to a background source produce strong absorption
features.  The most common extragalactic application of this technique
uses UV-bright quasars as the background sources \citep[for a recent
review see][]{bec03}.  Many species have been detected in absorption,
ranging from \hi\ and \heii, to \mbox{Fe\,{\sc ii}} and \mbox{Zn\,{\sc
ii}}, plus high-ionization ions such as \mbox{C\,{\sc iv}} and
\mbox{Si\,{\sc iv}} \citep[e.g.,][]{pro01}.

We examine the first ionization states of hydrogen and helium; they
are the most abundant elements in the IGM, and are responsible for
most of the UV opacity toward high-redshift quasars.  Specifically, we
will investigate the relative neutral fractions of hydrogen and
helium, which have first ionization potentials of 13.6 and 24.6~eV,
respectively.  Since the photoionization cross-sections for \hi\ and
\hei\ are peaked at threshold, their relative neutral fractions
constrain the effective slope of the ionizing background between 13.6
and 24.6~eV.

\begin{figure}
  \includegraphics[width=8.4cm]{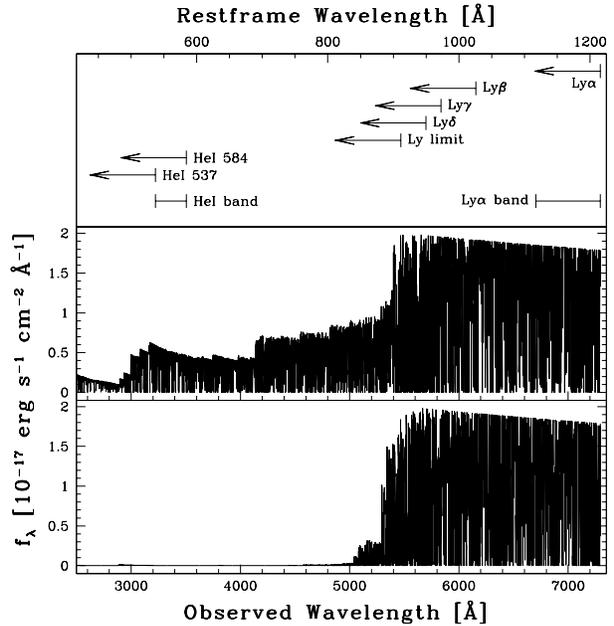}
  \caption{Simulated absorption spectra for $z=5$ quasars with
    AB$_{1450}=20$.  The top axis is labelled in rest-frame
    wavelength, and the bottom axis is labelled in the corresponding
    observed wavelength at $z=5$.  The top panel shows the wavelengths
    where the first four Lyman series line forests begin; it also
    shows where the the first two ground-state \hei\ transition
    opacities begin.  The \hei\ band is shown, as is the region of the
    corresponding \lya\ forest.  The middle panel shows a quasar
    spectrum processed by IGM absorption, along a line-of-sight with
    relatively little IGM opacity.  The bottom panel shows a quasar
    absorption spectrum along a typical line-of-sight.}
  \label{fig:scheme}
\end{figure}

The relative abundances of \hi\ and \hei\ constrain the ionizing
background---the challenge is to measure these abundances in the IGM
at high redshift.  Both atoms exhibit line and continuum absorption.
For \hi, the Lyman series lines begin at $1216~\ang$ and continue down
to 912~A, where continuous photoelectric absorption begins.  The
corresponding line transitions in \hei\ are from $1s^2$ to $1s2p$
(584~\AA), $1s3p$ (537~\AA), etc., down to the first ionization
threshold of helium, at 504~\AA.  These wavelengths are the rest-frame
values; the observed wavelengths of these transitions depend on the
redshift of the absorbing gas.  If the neutral gas is distributed in
many systems with a discrete redshift distribution, then the
associated absorption lines appearing in a quasar spectrum form a
`forest' of absorption features against the continuum.  Figure
\ref{fig:scheme} shows two example absorption spectra with the general
absorption regions labelled.

The absorption cross-sections of \hi\ and \hei\ are larger at the
center of a line (at the IGM temperature of $\sim 2\times
10^4~\kelvin$) than in the continuous absorption region.  Despite this
fact, the strongest absorption effect for the expected distribution of
neutral IGM gas (see Section \ref{sec:igmabs}) is from the cumulative
continuous absorption due to many absorbers at different redshifts,
named the `valley' by \citet{mol90}.  This is illustrated for \hi\
in Fig. \ref{fig:scheme}; once Lyman limit absorption begins
(shortward of $912(1+z_\rmn{Q})$~\AA, where $z_\rmn{Q}$ is the quasar
redshift), the forest typically gives way to almost complete
absorption.  In such cases it is very difficult to observe helium
forest lines because of the small residual continuum flux at short
wavelengths.  In rare cases the valley is not very pronounced, and the
continuum is strong enough that the \hei\ 584~\AA\ forest region may
be observed.  We call the region of the \hei\ 584~\AA\ forest red-ward
of \hei\ 537~\AA the `\hei\ band.'

The identification of absorption features in spectra like those in
Fig. \ref{fig:scheme} can be difficult, especially in low-resolution
spectra, except in the region red-ward of the Ly$\beta$ forest, where
the only possible absorption feature from hydrogen or helium is the
hydrogen \lya\ line.  (Trace metals produce additional absorption
features, though these can be discriminated based on linewidth or
because they have a doublet feature, e.g., \mbox{C\,{\sc iv}}.)  The
\hei\ band is the analogous region for \hei, where the \hei\ 584~\AA\
lines are the only source of helium opacity; however, this region also
may contain absorption due to all of the \hi\ Lyman transitions,
confusing the assignment of any individual line to \hei.  In
principle, unsaturated \hei\ lines could be identified because they
are twice as narrow as unsaturated \lya\ lines.  However, line
blending and the expected weakness of the \hei\ lines complicate a
practical implementation of that criterion, and spectroscopy of
sufficiently high resolution would require an extremely bright target.

Even after hydrogen reionization is complete at $z \simeq 6$, the
filamentary nature of the clumpy IGM leads to significant \lya\
opacity in regions with neutral hydrogen column density of $\nhi \sim
10^{14}~\cmmt$.  A line of sight through the universe pierces many of
these filaments at different redshifts, resulting in the \lya\ forest
described above.  Helium is expected to be singly reionized (24.6~eV
ionization threshold) at a similar redshift to hydrogen, for almost
any expected ionizing spectrum.  Consequently, absorption by neutral
helium at $z<6$ should be confined to the same filaments that give
rise to hydrogen absorption.\footnote{This is in contrast to doubly
ionized helium, which is distributed more diffusely (that is, the
absorption features traces smaller over-densities) than hydrogen
\protect\citep{kri01}.}  This justifies a one-to-one search for \hei\
584~\AA\ lines at the same redshift as lines identified in the `\lya\
band' region, where the \lya\ band is defined as the spectral region
where absorbers with a \hei\ 584~\AA\ line in the \hei\ band exhibit
\lya\ absorption (see Fig. \ref{fig:scheme}).

The application of absorption line studies of the IGM is limited by
the supply of suitable background sources, primarily quasars.  The
Sloan Digital Sky Survey\footnote{http://www.sdss.org} (SDSS) is a
large photometric and spectroscopic survey of the northern sky that
will identify and measure redshifts for almost all quasars brighter
than 20th magnitude over one quarter of the sky.  At the current
discovery rate of the survey, it will find approximately 1000 quasars
at $4<z5.2$, and about 200 at $4.8<z<5.2$ \citep{and01,fan01b}.  As we
will demonstrate (Section \ref{sec:qsosel}), rest-frame extreme
ultraviolet (EUV) absorption line study of $z \simeq 5$ quasars
requires a large sample of quasars, and could only be realized with
the large catalog of $z \simeq 5$ quasars that the SDSS will provide.

\section{Ionizing background}
\label{sec:ionback}

Next we consider the relationship between the
ionization state of IGM absorbers and the ionizing background.  The
quantity of interest for an absorbing system is the ratio of column
density of neutral helium, $\nhei$, to the column density of neutral
hydrogen, $\nhi$,
\begin{equation}
\eta \equiv \frac{\nhei}{\nhi}.
\end{equation}
As we will describe in Section \ref{sec:igmabs}, the abundance of
absorbers as a function of redshift and $\nhi$ is already observed.
Thus predictions for \hei\ absorption from these systems can be
easily made from calculations of the relative ionization states of
helium and hydrogen.

First we assume that the relative abundance of neutral helium to
neutral hydrogen in an absorbing system with high \hii\ and \heii\
fractions\footnote{We neglect the possible presence of \heiii\ in the
absorption systems.  Observations suggest that the second reionization
of helium did not occur until $z\simeq3$ \citep{kri01,the02}, and
\citet{wyi03b} predict that the full helium reionization happens over
a narrow redshift interval. These studies suggest that most of the
intergalactic helium was singly ionized at $z\simeq5$.} is described
by the optically-thin limit \citep[e.g.,][]{mir92},
\begin{equation}
\eta_\rmn{thin} \equiv \frac{\nhei}{\nhi} = \frac{n_\rmn{HeII}}{n_\rmn{HII}}
\frac{\alpha_\rmn{HeI}}{\alpha_\rmn{HI}}
\frac{\Gamma_\rmn{HI}}{\Gamma_\rmn{HeI}}
\simeq 0.083 \frac{\Gamma_\rmn{HI}}{\Gamma_\rmn{HeI}},
\end{equation}
where ${n_\rmn{He}}/{n_\rmn{H}} \simeq 0.083$ is the cosmic abundance
ratio of helium to hydrogen (assuming a mass fraction of $^4$He of
0.25, consistent with the combination of \textit{WMAP} CMB results
\citep{spe03} and big bang nucleosynthesis predictions \citep{bur01}),
${\alpha_\rmn{HeI}}/{\alpha_\rmn{HI}} = 1.0$ is the ratio of the
radiative recombination coefficients for \hei\ and \hi\ \citep{ost89},
and
\begin{equation}
\Gamma_\rmn{HI} = \int_{\nu_\rmn{HI}}^{\infty} \dd\nu 
\frac{4\pi J_\nu \sigma_\rmn{HI}(\nu)}{h \nu},
\end{equation}
and similarly for $\Gamma_\rmn{HeI}$.  The angle-averaged specific
intensity is $J_\nu$, $\sigma_i(\nu)$ is the photoionization
cross-section of species $i$, and the lower limit of the frequency
integral is $\nu_i$, the photon frequency at the ionization threshold
of species $i=${H,He}.

The most important absorbers for measuring $\eta$ are systems where
the observed equivalent widths of the \lya\ forest absorption lines
are roughly equal to the spectral resolution obtained in the \lya\
forest spectrum.  For a typical spectral resolution of 0.27~\AA\ (a
choice justified in Section \ref{sec:losqso_sa}), this corresponds to
$\nhi \simeq 3\times10^{14}~\cmmt$ at $z \sim 5$.  At such column
densities, the optical depth to photons at the hydrogen ionization
threshold is $\tau_\rmn{HI} \la 2\times10^{-3}$, thus the
optically-thin approximation should be valid.

However, optically thick absorbers can have a substantial impact on
the spectrum of the ionizing background, both through absorption and
emission.  \citet{haa96} find that the mean UV background is
attenuated by roughly a factor of 3 at both $\nu_\rmn{HI}$ and
$\nu_\rmn{HeI}$; additionally, much of the flux shortward of 228~\AA\
is absorbed by \heii\ and re-emitted at 304~\AA\ \citep[see][
fig. 5c]{haa96}.  These changes to the unabsorbed spectrum have
important implications for the \heiii\ abundance, but do not affect
the abundance ratio of \hei\ to \hi.

\begin{figure}
  \includegraphics[width=8.4cm]{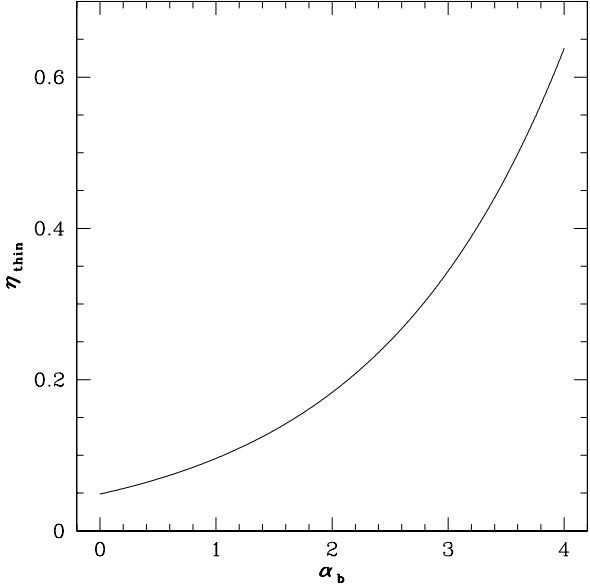}
  \caption{The ratio, $\eta_\rmn{thin}$, of the column density of
    neutral helium to the column density of neutral hydrogen, as a
    function of the spectral slope of the ionizing background,
    $\alpha_\rmn{b}$.  We assume that the absorbing systems are
    optically thin to ionizing radiation, highly ionized in hydrogen
    and highly singly-ionized in helium.  The effective power-law
    index of a normal stellar population is $\alpha_\rmn{b}=2.05$; the
    effective power law index of a population of very massive
    metal-free stars is $\alpha_\rmn{b}=0.28$.}
  \label{fig:optthin2}
\end{figure}

Because we are only concerned with the relative rate of neutral-helium
ionizations to hydrogen ionizations, we approximate the ionizing
background by a power law,
\begin{equation}
J_\nu \propto \nu^{-\alpha_\rmn{b}}.
\end{equation}
Using a continuous star-formation history with metallicity $Z=10^{-3}$
and a Salpeter IMF from 1 to 100~M$_\odot$ \citep{lei99}, the
effective power-law index is $\alpha_\rmn{b}=2.05$.  Also of interest
is the spectrum of very massive metal-free stars, which represent the
most extreme expectation of the stellar IMF at very high redshift:
these stars are well approximated by a blackbody spectrum with
temperature $T=10^5~\kelvin$ \citep*{bro01}.  Very massive metal-free
stars yield a particularly hard power-law index of
$\alpha_\rmn{b}=0.28$.  Quasars produce a spectrum with a power-law
shape matching their own EUV spectral slope; that value is not well
constrained at high redshift, but $\alpha_\rmn{b}=1.6$ may be a
reasonable guess (see Section \ref{sec:qsospec}).  Finally, x-rays
radiated from early stars have a near-zero spectral slope
\citep{oh01}, but the effect of secondary ionizations should produce
an effective slope slightly greater than zero (S.P. Oh, private
communication).  Figure \ref{fig:optthin2} shows the value of
$\eta_\rmn{thin}$ as a function of $\alpha_\rmn{b}$.

\section{IGM absorption model}
\label{sec:igmabs}

Our model for the \hi\ absorption in the IGM is empirical, based on
the data obtained in many quasar absorption line studies. Ongoing and
planned observations are likely to constrain its parameters better in
the future.

Absorption by \hi\ is quantified by discretizing the absorption into
individual absorbing systems, identified by their redshift, $z$, and
neutral hydrogen column density, $\nhi$.  The distribution of \hi\
systems may be described reasonably well by power laws in both
redshift and column density.  We define \hi\ systems with
$\nhi<1.6\times10^{17}~\cmmt$ as \lya\ forest systems, and systems
with $\nhi\ge1.6\times10^{17}~\cmmt$ as Lyman Limit Systems (LLSs).
The distribution of \hi\ systems in redshift and column density are
parametrized by three constants: $A$, $\gamma$, and $s$,  
\begin{equation}
\frac{\dd N(z)}{\dd z} = A (1+z)^\gamma,
\end{equation}
\begin{equation}
f(\nhi) \propto \nhi^{-s}.
\end{equation}
Here $N(z)$ is the total number of absorbers along a line of sight to
redshift $z$, and $f(\nhi)$ is the number of absorbing systems per
unit \hi\ column density.

We chose $s=1.5$ for the power law index of the $\nhi$ distribution
for all values of $\nhi$, and assume that the absorber population
extends up to $\nhi=10^{22}~\cmmt$ \citep{sto00}.  For the \lya\
forest, we adopt the redshift evolution determined by \textit{HST}
observations at low redshifts (\citealt{dob02}, cf. \citealt{wey98})
and from ground-based surveys at high redshifts \citep{bec94}.  Those
surveys covered the \hi\ column density range of $10^{14}~\cmmt \leq
\nhi \leq 1.6\times10^{17}~\cmmt$ (i.e., from a rest-frame equivalent
width of 0.24~\AA\ up to the LLSs).  We extend their results to $\nhi
= 10^{13}~\cmmt$ by scaling up the number density using the $\nhi$
distribution quoted above, to give
\begin{equation}
\begin{array}{c}
A=106,\quad\gamma=0.65,\quad\mathrm{for }~z<2.4, \\
A=29,\quad\gamma=1.7,\quad\mathrm{for }~z\ge2.4.
\end{array}
\end{equation}
The transition redshift, $z=2.4$, was chosen to match the low-redshift
and high-redshift fits smoothly; this is higher than the traditional
value of about 1.5 \citep[e.g.,][]{wey98}, but is a consequence of the
upward revision of $\gamma$ by \citet{dob02} compared to
\citet{wey98}; given the large uncertainties in the values of $A$ and
$\gamma$, this is not alarming.

For the LLSs, we adopt $A=0.2$ and $\gamma=1.5$, which are
consistent with the results of \citet{sto94} and \citet{ste95}.  We
note that \citet{ste95} do not present an analysis of their full
sample of LLSs because they exclude absorption systems within
$5000~\kms$ of the quasar, yet their results are consistent with
the result of \citet{sto94} for all LLSs.  These parameter choices provide
a relatively smooth intersection between the abundance of LLSs and
\lya\ forest systems at $z \ge 1$, assuming the column density
distribution described above.  The covariance of the uncertainties in
$A$ and $\gamma$ implies that the mean number of LLSs toward a
high-redshift quasar is constrained, but the LLS distribution with
redshift is uncertain: the 1-$\sigma$ uncertainty on $\gamma$ is about
0.4 \citep{sto94,ste95}.

The absorber distributions described above can be converted into a
mean comoving \hi\ number density using
\begin{equation}
\left< n^\rmn{c}_\rmn{HI}(z) \right> = A (1+z)^\gamma 
\frac{H_0 E(z)}{c (1+z)^2} \left< \nhi \right>,
\end{equation}
where $H_0$ is the Hubble constant, $\left< \nhi \right>$ is the mean \hi\
column density of the absorber population, and
\begin{equation}
E(z) = \left[ \Omega_\rmn{m} (1+z)^3 + \Omega_\Lambda + 
(1-\Omega_\rmn{m}-\Omega_\Lambda)(1+z)^2 \right]^{1/2}.
\end{equation}
For a universe with density parameters $\Omega_\rmn{m}=0.3$ in matter,
$\Omega_\Lambda=0.7$ in a cosmological constant, and
$\Omega_\rmn{b}h^2=0.02$ in baryons, and a Hubble constant $h\equiv
H_0/(100~\mathrm{km~s^{-1} Mpc^{-1}})=0.7$, this gives an \hi\ neutral
fraction at $z \ga 4$ of,
\begin{equation}
\left< x_\rmn{HI}(z) \right> \simeq 9.8\times10^{-3} A (1+z)^{\gamma-1/2} 
\frac{\left< \nhi \right>}{4\times10^{19}~\cmmt}.
\end{equation}

For LLSs (which have $\left< \nhi \right>=4\times10^{19}~\cmmt$),
$A=0.2$ and $\gamma=1.5$, thus $\left< x_\rmn{HI}(z=6) \right>=0.01$.
This is consistent with the limit derived by \citet{fan02} from \hi\
absorption toward a $z=6.28$ quasar.

We assume that there is no correlation between absorbing systems.
There is observational evidence for clustering of the \lya\ forest
lines \citep[e.g.,][and references therein]{lis00,dob02}, but our
assumption is conservative for the purposes of this paper: we will
show that to study \hei\ absorption, we need quasars with fewer than
average absorbers along the line of sight.  Clustering of absorbers
would skew the `bad' quasar targets (i.e., those with many strong
absorbers) worse, and skew the `good' quasar targets better.  This is
not an important effect for low-$\nhi$ absorbers, since they are very
numerous, but could be a significant effect for LLSs if they are
clustered (note, however, that \citealt*{sar89} showed that LLSs
followed Poisson statistics in their sample of 37 absorbers).  The
proximity effect generates an absorber deficit of $\sim 30$ per cent
within $4~h^{-1}~\mpc$ of the quasar, or about 200~\AA\ observed from
the quasar \lya\ line \citep{sco00}.  We also ignored this relatively
small effect.

In addition to the absorber redshift and column-density distributions,
our IGM model must also describe their absorption properties.  We used
the photoionization cross-sections given by \citet{ost89} for \hi\ and
\citet{ver96} for \hei.  Our model treats the first 11 line
transitions from the ground state (which all atoms are assumed to
occupy), using data from the NIST Atomic Spectra Database
v2.0.\footnote{http://physics.nist.gov/cgi-bin/AtData/main\_asd} The
line profiles were modelled as Doppler cores of width $b=26~\kms$
\citep{kim97} with damping wings outside of the core
\citep[e.g.,][]{pee93}.\footnote{In Section \ref{sec:losqso_sa} we
assume instrumental resolutions considerably larger than $26~\kms$, so
the choice of exactly that value and our use of a single $b$ value for
all absorbers rather than the actual distribution should have a
negligible effect on our results.}

Our absorption model includes only hydrogen and helium; metal lines
are not considered here.  There are not likely to be many metal line
systems in the quasars used to measure $\eta$ because metal lines are
associated with \hi\ absorbers of high column density, and, due to the
selection techniques employed (see Section \ref{sec:losqso}), quasars
with strong absorbing systems are unsuitable for measuring $\eta$.
Moreover, the strongest metal lines that could pollute the \lya\ band,
such as the \mbox{C\,{\sc iv}} and \mbox{Mg\,{\sc ii}} doublets, show
only a few absorbers per unit redshift, compared to of order 100 \lya\
absorbers in the same redshift interval.  In the \hei\ band, metal
lines are again far less numerous than \hi\ lines, and thus were
ignored.

One complication to the application of absorber statistics determined
from other quasar samples to the population of quasars discovered by
SDSS is the quasar selection techniques employed.  The SDSS quasar
selection primarily uses observed optical colors to generate a
well-defined selection function \citep{fan99}; by contrast, the
quasars studied in absorption line surveys were culled from
heterogeneous catalogs \citep[e.g.,][]{hew87}, and include radio-,
x-ray-, and emission-line-selected quasars in addition to
color-selected quasars.  An analysis of the systematic errors
introduced by using absorption lines in quasars discovered by several
different techniques to predict absorption patterns in quasars
selected by another technique is outside the scope of this paper; we
simply analyze the colors of the quasars most important to this study
to ensure they would meet the quasar color-selection criteria of SDSS
(see Section \ref{sec:qsosel}).

\section{Model quasar intrinsic spectrum}
\label{sec:qsospec}

In order to simulate realistic observations of IGM absorption of
background quasars, we need to make assumptions about the intrinsic
properties of the quasar UV spectrum.  The SDSS will provide a large
catalog of high-redshift quasars, so we attempt to model the typical
properties expected of SDSS quasars based on the results of the survey so
far.

In a sample drawn from 182 square degrees of the SDSS, \citet{fan01a}
presented 18 quasars with $z\ge4$ and $i^*<20$, where $i^*$
is the preliminary SDSS determination of the quasar $i'$ magnitude.  We
assume the quasars presented are a fair sample of the final SDSS results.

For each quasar in their sample, \citet{fan01a} compute the properties of
the quasar near-UV (NUV) continuum, assuming a power law form,
$f_\nu\propto\nu^{-\alpha_\rmn{NUV}}$.  They set the normalization to
the quasar continuum at the observed wavelength corresponding to
rest-frame 1450~\AA.  They then convert that specific flux into a magnitude on
the AB system, 
\begin{equation}
\mathrm{AB}_{1450} = -2.5 \log_{10}
[f_\nu(\lambda_\rmn{rest}=1450~\ang)] - 48.6,
\end{equation}
where $f_\nu(\lambda_\rmn{rest}=1450)$ is the specific flux in $\fnu$
at rest-frame 1450~\AA.  Typically AB$_{1450} \simeq i^*$ or $z^*$,
depending on quasar redshift. \citet{fan01a} also estimate
$\alpha_\rmn{NUV}$ for each quasar; the mean for quasars with $z\ge4$
is 0.6, with substantial uncertainty in the slope of any individual
quasar.  However, \citet{tel02} found a break in the UV slope of their
quasar composite spectrum near the wavelength of \lya; the measured
slope became softer blue-ward of \lya.  This confirmed earlier work by
\citet{zhe97}.  The \citet{tel02} sample is comprised almost entirely
of quasars with $z<2.5$; for radio-quiet quasars they find a mean EUV
slope of about 1.6, with a break to a shallower NUV slope at about
1250~\AA.  \citet{tel02} compare their radio-quiet sample to the
radio-quiet SDSS quasar sample of \citet{van01}, and find an evolution
toward harder NUV slope with samples at higher redshift; EUV
comparison between the samples is extremely difficult due to the
presence of strong IGM absorption in the high-redshift quasars.

Given the available observational evidence, we adopt a conservative
model for the EUV properties of SDSS quasars: AB$_{1450}=20$ with a NUV
slope of 0.6 between rest-frame 1450 and 1250~\AA\ and an EUV slope
shortward of 1250~\AA\ given by $\alpha_\rmn{EUV}=1.6$.

\section{Lines of sight toward SDSS quasars}
\label{sec:losqso}

Many $z\sim5$ quasars will be discovered by SDSS; in this Section we
evaluate how good the best quasar target will be.  We then simulate
observations of the best quasar target and demonstrate our ability to
constrain the spectrum of the ionizing background from such data.

Extrapolating from the results of the color-selected sample of
\citet{fan01a}, SDSS will discover $\sim 1000$ quasars with $4<z<5.2$.
We assume the SDSS quasar selection function does not depend much on
redshift over the range $4<z<5.2$; this is a conservative estimate for
our purposes because the detection probability is higher for quasars
with $4.7<z<5.2$ than for other $z>4$ quasars \citep{fan01a}.
\citet{fan01b} fit the redshift dependence of the high redshift quasar
spatial density with $\rho \propto 10^{-0.5z}$.  Thus about 200 of the
high-redshift quasars will fall within the range $4.8<z<5.2$.

We would like to know how many of these quasars have spectra like the
middle panel of Fig. \ref{fig:scheme}, which would be useful for
measuring \hei\ lines, and how many have spectra like the bottom panel
of Fig. \ref{fig:scheme}, which would not be a suitable target for
measuring \hei\ lines.  We quantify this by determining the
distribution of quasar sightline `suitability,' as measured by the
flux of the quasar in the \hei\ band.  From that distribution and the
expected number of SDSS quasars, we determined what the most suitable
quasar discovered by the SDSS will be.  Though for some applications
an analytic approach to the subject is suitable \citep{zuo93}, we
require a Monte Carlo simulation approach \citep{mol90,jak98} for this
study.

\subsection{Absorber Monte Carlo simulations}
\label{sec:losqso_mc}

We started from the distribution of \hi\ absorbing systems in redshift
and column density described in Section \ref{sec:igmabs}.  For a
quasar at a given redshift $z_\rmn{Q}$, we used Poisson statistics to
generate the number of low-$z$ \lya\ forest, high-$z$ \lya\ forest and
LLS absorbers along the line of sight.  Once we generated the number
of absorbers of each type, we assigned redshifts and column densities
drawn from the absorber model described in Section \ref{sec:igmabs}.
This list of absorbers was then passed through a routine to calculate
the optical depth to \hi\ at every sampled wavelength
value.\footnote{Wavelength sampling ranges from 0.08 to 0.28~\AA,
chosen so that there are three samples per Doppler FWHM of the
spectral lines of interest (\hei\ lines are twice as narrow as \hi\
lines); all results were checked for sensitivity to spectral
resolution and are converged at this resolution choice.}  In each
simulated spectrum, the optical depth due to helium was stored
separately from hydrogen; given the universal optically thin ratio of
column densities, $\eta_\rmn{thin}$ derived in Section
\ref{sec:ionback}, the helium optical depth scales simply as
$\eta_\rmn{thin}$.

\begin{figure}
  \includegraphics[width=8.4cm]{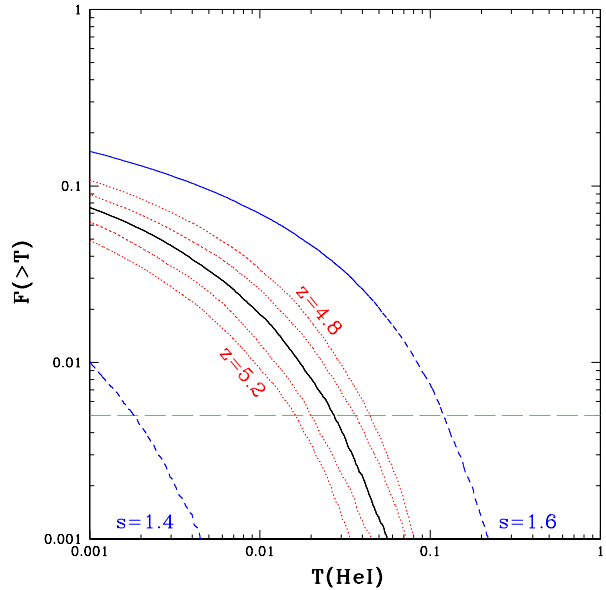}
  \caption{Cumulative histogram of the \hei-band transmission along
    quasar lines-of-sight, as a function of redshift and absorber
    model.  The solid curve shows the fraction of $z=5$ quasars that
    have a mean transmission in the \hei\ band greater than a given
    mean transmission value $T(\hei)$, assuming $s=1.5$ in the
    distribution of absorber column densities (Section
    \ref{sec:igmabs}).  The lowest dashed curves are for $s=1.4$ and
    $s=1.6$, respectively, assuming $z=5$ quasars.  The two dotted
    curves closely above the solid curve are for $z=4.9,4.8$, and the
    two dotted curves closely below the solid curve are for
    $z=5.1,5.2$, all assuming $s=1.5$.  The horizontal dashed line at
    1/200 intersects the curves at the expected $T(\hei)$ of the best
    quasar in a sample of 200 quasars.}
  \label{fig:suitb}
\end{figure}

For each LOS, the \hi\ optical depth data were converted into the IGM
transmission as a function of wavelength.  We integrated the IGM
transmission over the \hei\ band to derive the mean \hei-band
transmission along each LOS, $T(\hei)$.  Figure \ref{fig:suitb} shows
the cumulative fraction of quasars with $T(\hei)$ greater than a given
value.  For simplicity, we will assume the expected 200 SDSS quasars
with $4.8<z_\rmn{Q}<5.2$ all fall at $z_\rmn{Q}=5$.  Then from Fig.
\ref{fig:suitb} we can read off the value of $T(\hei)$ corresponding
to a fraction of $1/200$; this is the largest expected $T(\hei)$ value
of the lines-of-sight toward the SDSS quasars.  The result is
$T(\hei)=0.03$.  Note that if we change the assumed redshift of the
quasars by $\Delta z=0.2$ (the dotted lines in Fig.  \ref{fig:suitb}),
the expected maximum value of $T(\hei)$ for 200 quasars changes by a
factor of about 2.  When the final SDSS quasar catalog is constructed,
we may use the real distribution of quasar redshifts together with our
simulation machinery to generate a $T(\hei)$ histogram that depends
only on the IGM absorber model.  Thus measurements of the observed
distribution of $T(\hei)$ can be used to constrain a combination of
the properties of the \hi\ component of the IGM and the intrinsic
quasar spectral shape.

Small changes to the absorber-model parameter $s$, the slope of the
column-density distribution, have a strong effect on the expected
maximum value of $T(\hei)$ (as pointed out by \citealt{mol90}, and
illustrated by the dashed lines in Fig. \ref{fig:suitb}).  In the next
subsections we discuss what could be learned about the $z\simeq5$
ionizing background under the assumption that SDSS will discover a
quasar with properties given in Section \ref{sec:qsospec} along a LOS
with $T(\hei)=0.03$, our `expected' value for the best SDSS quasar, or
$T(\hei)=0.1$, which we consider to be a reasonable `optimistic' value
for the best SDSS quasar.

\subsection{Quasar absorption line spectrum simulations}
\label{sec:losqso_sa}

Based on the results of the previous subsection, we analyzed
lines-of-sight with $T(\hei)=0.03$ and $0.1$, our `expected' and
`optimistic' values for the best SDSS quasar LOS.  For each
observational realization of a quasar along the simulated LOS, a value
of $\eta$ was assumed.  Then the optical depth data were used to
construct the IGM transmission as a function of wavelength.  This was
multiplied by the intrinsic quasar spectrum from the model described
in Section \ref{sec:qsospec} to generate the observed spectrum.  The
Galactic extinction curve was taken from \citet{sch88}.  Atmospheric
extinction was applied based on the sky transparency from Mauna Kea
(the location of the Keck
Observatory).\footnote{http://www2.keck.hawaii.edu/inst/lris/atm\_trans.html}
Our modelling of the telescope and instrument parameters included slit
losses, system throughput, instrumental resolution, pixel sampling,
spatial extraction of the spectra, and detector noise.  Slit losses
and spatial extraction both incorporate an assumed value for the
seeing.  A mean emission spectrum of the
sky,\footnote{http://www.cfht.hawaii.edu/Instruments/ObserverManual/
chapter5.html; note the sky will darken slightly as we approach solar
minimum in 2008.} measured at Mauna Kea, was also passed through the
telescope/instrument model.

Our model observations were performed assuming a Galactic extinction
value of $A_{u'}=0.15$ toward the target quasar, typical for SDSS
quasars \citep{sch02}.  We assumed the observations were made at a
constant value of 1.2 airmasses and a constant seeing of 0.7~arcsec.

Our simulated observations of the \hi\ \lya\ forest were based on the
properties of the Keck II 10-meter telescope and the Echellette
Spectrograph and Imager (ESI; \citealt{she02}), an intermediate
resolution optical spectrograph, with ESI in echelle mode
($R\sim4000$).  We assumed a 1~arcsec slit, with $75~\kms$ resolution
and $11.5~\kms\mathrm{pixel}^{-1}$.  The spatial extraction window was
0.77~arcsec; the spatial pixel scale was
$0.154~\mathrm{arcsec~pixel}^{-1}$.  We assumed no dark current, and a
readnoise of 2.1 counts per pixel (the gain was 1.3 e$^{-}$ per
count).  We assumed each \hi\ \lya\ forest spectrum consisted of 40
co-added 1000~sec observations, for a total exposure time of 11 hours.
This is similar to the exposure times of Keck observations of the
highest-redshift SDSS quasars \citep{whi03}.

We assumed that the \hei\ band was observed with the Keck I 10-meter
telescope and the Low-Resolution Imaging Spectrometer (LRIS;
\citealt{oke95}), using the blue channel of LRIS (LRIS-B;
\citealt{mcc98}) with the 1200 line grism.  This configuration with a
0.7~arcsec slit delivers a resolution of $1.30~\ang$, sampled at
$0.24~\ang\,\mathrm{pixel}^{-1}$.  The spatial extraction window was
1.08~arcsec, about 1.5 times the seeing; the spatial pixel scale was
$0.2151~\mathrm{arcsec~pixel}^{-1}$.  We assumed no dark current, and
a readnoise of 2.5 counts per pixel (the gain was 1.6 e$^{-}$ per
count).  Because readnoise otherwise made a non-negligible
contribution to the noise, we binned the data by 2 pixels in the
spatial and spectral directions.  We assumed each \hei\ 584~\AA\
forest spectrum consisted of 72 co-added 2000~sec observations, for a
total exposure time of 40 hours.  This is substantially more
observation time than is typically devoted to any one object with
Keck.  However, the total observation time devoted to study \heii\
absorption in the $z=2.885$ quasar HE2347-4342 was almost 150 hours
with the \textit{FUSE} satellite.

For each simulated exposure, we computed the expected signal, expected
sky, and expected detector noise.  The observed counts in each
spectral pixel due to source, sky and detector were drawn from Poisson
distributions with the means set to the expected counts, and the three
contributions were summed.  Then the expected sky and detector noise
were subtracted (assuming photon-noise limited sky subtraction).
The simulated exposures were then summed to create the final simulated
observation.

\subsection{Quasar absorption line spectrum analysis}
\label{sec:losqso_anal}

\begin{figure}
  \includegraphics[width=8.4cm]{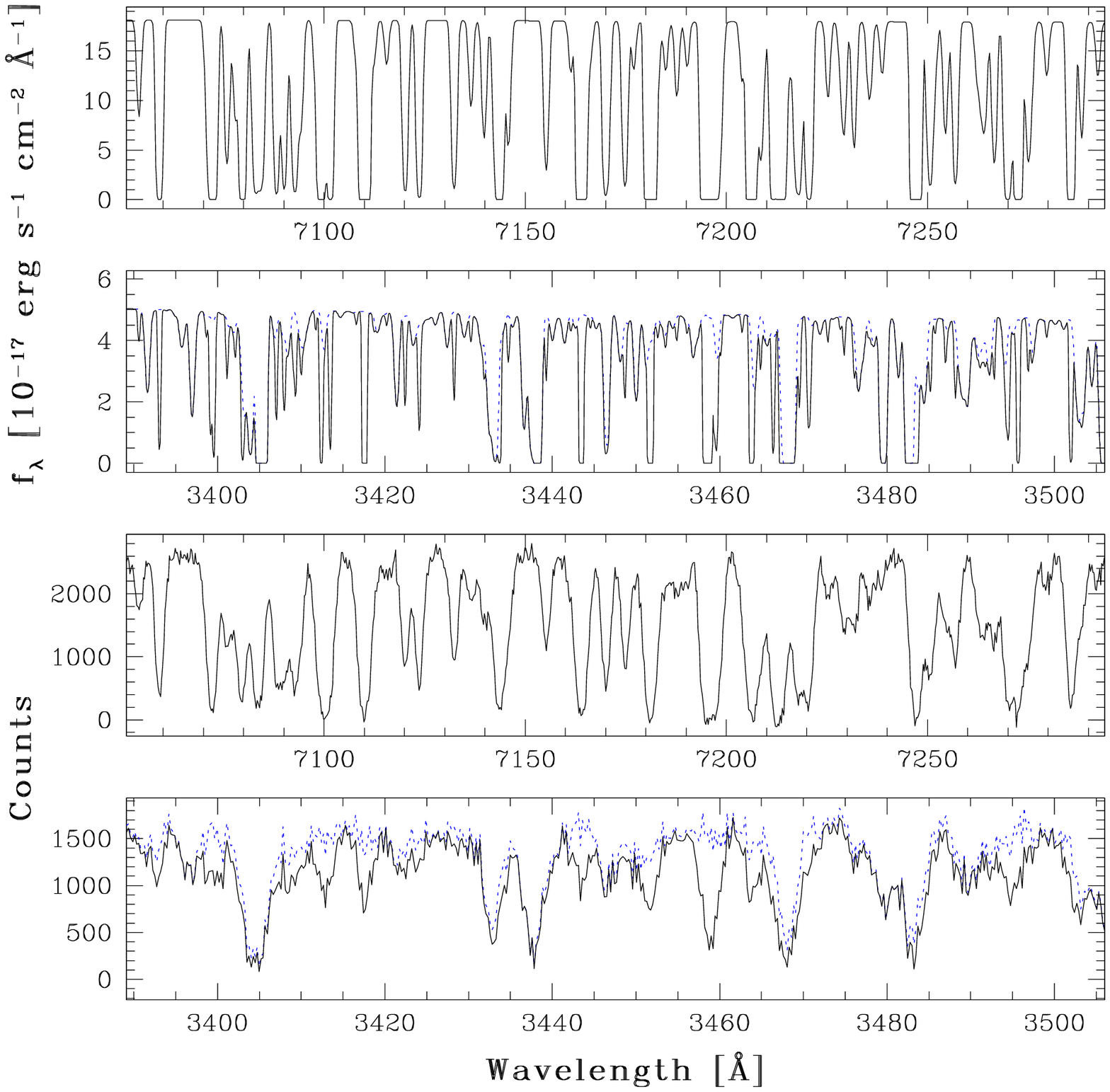}
  \caption{A simulated absorption spectrum for a $z=5$ quasar with
    AB$_{1450}=20$.  The top panel zooms in on the $z=4.8$--5 \hi\
    \lya\ forest of the spectrum from the middle panel of
    Fig. \ref{fig:scheme}.  The second panel shows the corresponding
    $z=4.8$--5 \hei\ 584~\AA\ forest region of the same spectrum; the
    dotted curve is the absorption from \hi\ only, and the solid curve
    assumes $\etathin=0.5$ for illustration.  The lower two panels
    show mock observations of the quasar, using the observation model
    described in Section \ref{sec:losqso_sa}.}
  \label{fig:illus}
\end{figure}

Our goal is to estimate the ionization state of the IGM as measured by
the relative abundance of \hei\ to \hi, $\eta$.  Our analysis
technique is to make use of the cross-correlation of the \hi\ \lya\
forest spectrum and the \hei-band spectrum, which depends on $\eta$
as illustrated in Fig. \ref{fig:illus}.

\begin{figure}
  \includegraphics[width=8.4cm]{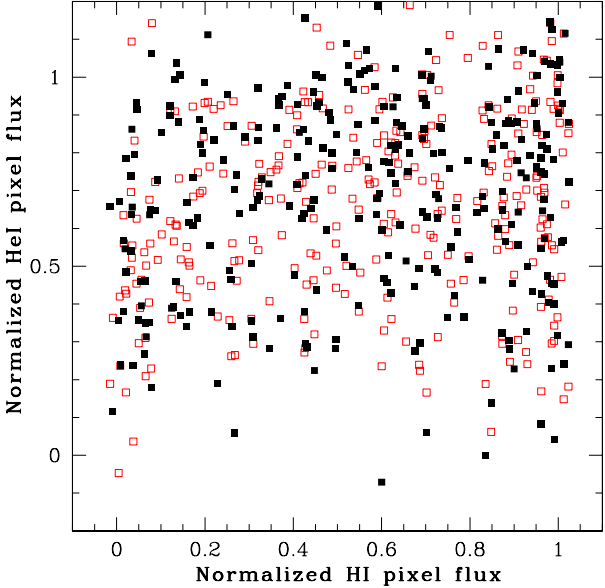}
  \caption{An example of the pixel-by-pixel cross-correlation of the
    absorption in the \hei\ band and the \hi\ \lya\ band.  The solid
    squares represent continuum-normalized flux pairs from an
    observation of a quasar LOS with $\eta=0.06$; the open squares are
    from an observation of the same LOS but assuming $\eta=0.20$.}
  \label{fig:p1}
\end{figure}

Each pixel in the \hei-band spectrum was matched up with the \lya-band
spectrum pixel whose wavelength is closest to
$\lambda_\rmn{HI}=(1216/584)\lambda_\rmn{HeI}$, where
$\lambda_\rmn{HeI}$ is the wavelength of the \hei-band pixel. Thus an
absorber at a given redshift will exhibit \hei\ 584~\AA\ absorption
and \lya\ absorption in matched pixels.  Figure \ref{fig:p1} shows a
plot of the fluxes, normalized to the local quasar continuum, of pixel
pairs matched in this manner, for two values of $\eta$.  If \hei\
584~\AA\ absorption and \lya\ absorption were the only features in the
spectra, and the observations were perfect, then the points would fall
along a tight locus described primarily by $\eta$.  In regions of
little absorption, both continuum-normalized pixel fluxes would be
near one: regions of strong absorption would have normalized \hi\
pixel fluxes near zero and normalized \hei\ pixel fluxes set by
$\eta$.  However, the presence of other lines and the effects of
instrumental smoothing and observational noise scatter the points.
There is little scatter in the \hi-pixel-flux direction, both because
these simulated observations have a good signal-to-noise ratio, and
because there are no absorption features in the \lya\ band besides
\lya.  There is substantial scatter in the \hei-pixel-flux direction,
though, because of lower signal-to-noise in the observations, and also
the presence of \lya\ absorption (and, to a lesser extent, absorption
due to the other Lyman series lines) from low-redshift absorbers.
Despite the scatter, one can easily see that, at low values for the
normalized \hi\ pixel flux, the points Fig. \ref{fig:p1} for the
$\eta=0.06$ case (solid squares) lie higher (less \hei\ absorption)
than the points for the $\eta=0.20$ case (open squares).

In the \hei-band spectrum, we would like to know what the quasar
effective continuum level is after accounting for bound-free
absorption by higher-redshift absorbing systems.  For a $z_\rmn{Q}=5$
quasar, any strong absorber between $z=2.53$ and $z=2.84$ will cause a
change in the effective continuum in the \hei\ band, due to the \hi\
bound-free absorption edge at $912(1+z)~\ang$.  Consequently, the \hi\
\lya\ forest spectrum should be searched for strong \lya\ lines over
the range of 4292 to 4669~\AA\ (for $z_\rmn{Q}=5$).  We assumed for
our analysis that the effective continuum can be accurately estimated.
In the analysis of real observations, one could apply the same
technique used to estimate the effective continuum to our mock
observations, and thus analyze the real observations in an unbiased
way.

\begin{figure}
  \includegraphics[width=8.4cm]{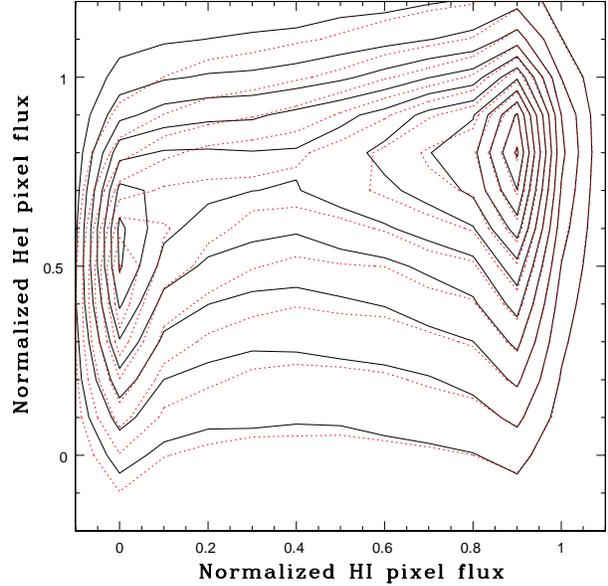}
  \caption{Contour plot of the ensemble continuum-normalized fluxes of
    pairs of corresponding \hei-band and \lya-band pixels (see Section
    \ref{sec:losqso_anal}).  The solid contours assume $\eta=0.06$,
    and the dotted contours are for $\eta=0.20$.  Pixel pairs with
    strong \hi\ absorption have \hi\ pixel fluxes near zero; for those
    pairs increasing $\eta$ shifts the \hei\ pixels to smaller fluxes,
    due to increased \hei\ absorption associated with the \hi\
    absorption.}
  \label{fig:bins}
\end{figure}

Using our large Monte Carlo library of simulated quasar LOSs, we
generated the distribution of pixel pair fluxes, as a function of
$\eta$, for multiple observational realizations of a quasar along each
LOS.  Figure \ref{fig:bins} shows contours that represent the
likelihood of finding a given pair of continuum-normalized fluxes for
$\eta=0.06$ and $\eta=0.20$.  This distribution of mock-observational
results were then used to analyze an individual observation of a
quasar LOS: we generated the matched pair data as shown in
Fig. \ref{fig:p1}, then summed the value of the likelihood at that
point over all the pixel pairs with continuum-normalized \hi\ pixel
fluxes less than 0.5 (cf. Fig.  \ref{fig:bins}), for each of the two
values of $\eta$ under consideration.  We termed each sum of the
likelihood values (which depend on $\eta$) $L(\eta)$.  The $L(\eta)$
value is larger when a set of pairs matches up well with the contours
of Fig. \ref{fig:bins} for $\eta$, and smaller when there is not a
good match.  Then we formed the ratio $R\equiv L(0.06)/L(0.20)$, which
we find is better for recovering our input value of $\eta$ than, for
example, $L(0.06)-L(0.20)$.  We have found that binning the \hei-band
spectra by 2 pixels (in addition to the $2\times2$ on-chip binning)
reduces the scatter in $R$.

\begin{figure}
  \includegraphics[width=8.4cm]{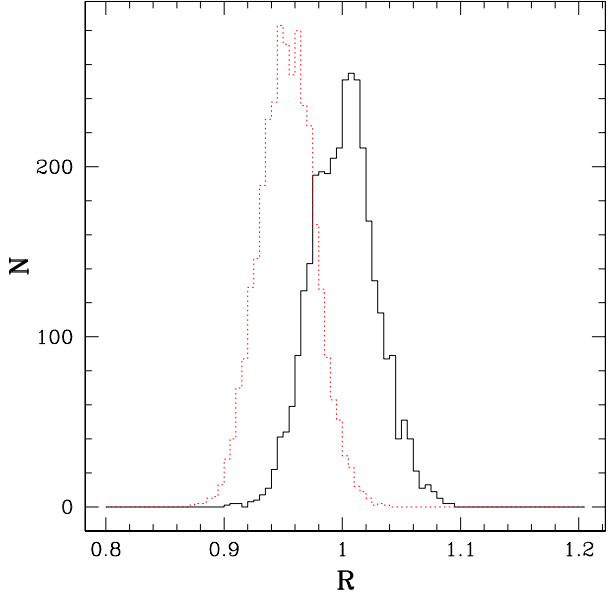}
  \caption{Histograms of $R$ (see Section \ref{sec:losqso_anal}) for
    each of our two input values of $\eta$, 0.06 and 0.20, assuming
    observation of a quasar along a sightline with $T(\hei)=0.03$.
    The solid (dotted) line shows the $R$ histogram for analysis of
    quasar LOSs with $\eta=0.06$ ($\eta=0.20$).}
  \label{fig:compexp5}
\end{figure}

\begin{figure}
  \includegraphics[width=8.4cm]{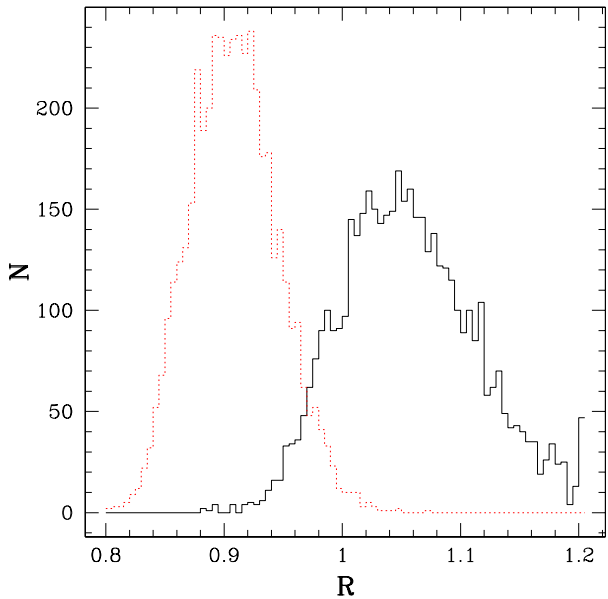}
  \caption{Same as Fig. \ref{fig:compexp5} but for an LOS with
    $T(\hei)=0.1$.}
  \label{fig:compopt5}
\end{figure}

Figure \ref{fig:compexp5} shows histograms of $R$ for each of our two
input values of $\eta$, assuming observations described in Section
\ref{sec:losqso_sa} of a quasar along a sightline with $T(\hei)=0.03$,
our {\it expected} value for the best SDSS LOS toward a $z\simeq5$
quasar (see Section \ref{sec:losqso_mc}).  The histograms are clearly
separated, though they do show some overlap.  Figure
\ref{fig:compopt5} is the same as Fig. \ref{fig:compexp5}, but for
observations along a LOS with $T(\hei)=0.1$, our {\it optimistic}
value for the best quasar LOS in the SDSS.  In this case the
histograms are much more clearly separated, illustrating the
importance of discovering a very good quasar LOS.

\begin{figure}
  \includegraphics[width=8.4cm]{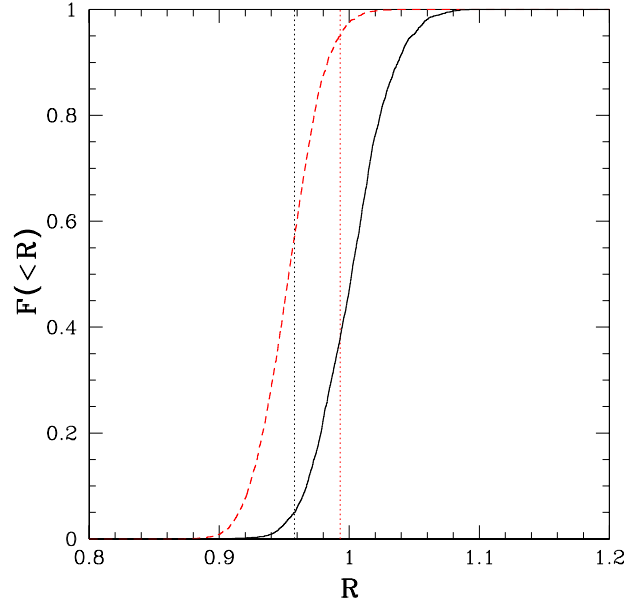}
  \caption{Normalized cumulative histograms of $R$ from Fig.
    \ref{fig:compexp5}; 95 per cent of the LOSs have $R(0.06)>0.958$,
    the position of the left vertical line, and 95 per cent of the
    LOSs have $R(0.20)<0.993$, the position of the right vertical
    line.  Thus if the intrinsic $\eta$ value along a LOS is either
    0.06 or 0.20, one observation has about a 60 per cent chance of
    rejecting the other value.}
  \label{fig:cumcompexp}
\end{figure}

\begin{figure}
  \includegraphics[width=8.4cm]{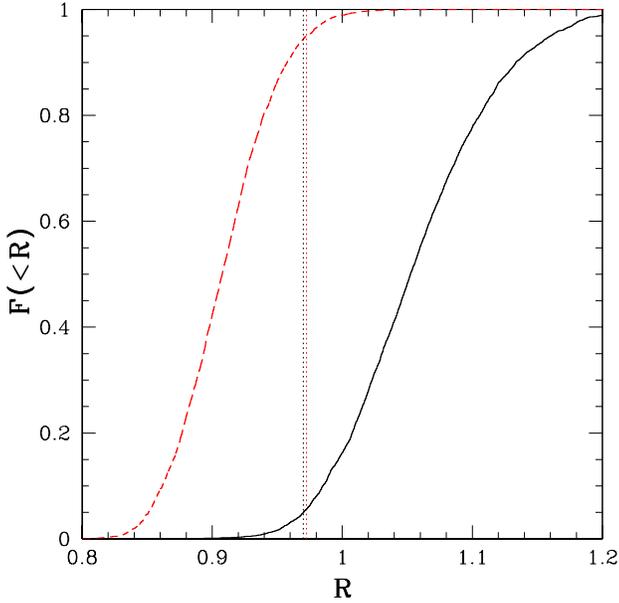}
  \caption{Normalized cumulative histograms of $R$ from Fig.
    \ref{fig:compopt5}; 95 per cent of the LOSs have $R(0.06)>0.970$,
    the position of the left vertical line, and 95 per cent of the
    LOSs have $R(0.20)<0.972$, the position of the right vertical
    line.  Thus if the intrinsic $\eta$ value along a LOS is either
    0.06 or 0.20, one observation has about a 95 per cent chance of
    rejecting the other value.}
  \label{fig:cumcompopt}
\end{figure}

Figures \ref{fig:cumcompexp} and \ref{fig:cumcompopt} cumulate (and
normalize) the histograms of Figs. \ref{fig:compexp5} and
\ref{fig:compopt5}.  They illustrate that, for our model observations
toward one $T(\hei)=0.03$ LOS, if the true value of $\eta$ is 0.06,
then we have almost a 60 per cent chance of making an observation that
will reject $\eta=0.20$ at the 95 per cent confidence level.
Conversely, if the true value of $\eta$ is 0.20, then we have a 60 per
cent chance of making an observation that will reject $\eta=0.06$ at
the 95 per cent confidence level.  The situation along a $T(\hei)=0.1$
LOS is much more optimistic: if the true value of $\eta$ is either one
of our model choices, we will reject the other value of $\eta$ about
95 per cent of the time at the 95 per cent confidence level.

Using the observations we propose, it will be difficult to estimate a
precise value of $\eta$ from the data.  If the true value of $\eta$
is, for example, 0.11, then we will only be able to reject very
extreme values of $\eta$.  However, if the ionizing background is
either hard or soft, it may well be possible to reject the other
hypothesis.  For example, if we measured an $R$ value of 1, we could
be confident that the ionizing background at $z\simeq5$ is not
dominated by \popii\ stars.

\section{Observational quasar selection techniques}
\label{sec:qsosel}

In previous Sections we estimated the likely properties of the best
SDSS quasar for the measurement of \hei\ absorption features, and what
we could learn about the $z\simeq5$ ionizing background from analysis
of such a quasar.  In this Section we describe the final aspect of
practical implementation, how to pick out the best SDSS quasar from
the expected sample of 200.

We would like to select the $z\simeq5$ SDSS quasar with the highest flux
in the \hei\ band.  So far we have primarily discussed $T(\hei)$, the
IGM-transmitted fraction in the \hei\ band; however, this was always
under the assumption of the intrinsic quasar model described in Section
\ref{sec:qsospec}.  The real SDSS quasar sample will certainly include
quasars with a range of values for AB$_{1450}$, and these quasars may have a
range of EUV spectral shapes.  Due to these distributions, the quasar
with the largest \hei-band flux may not be the quasar with the largest
$T(\hei)$; we made conservative assumptions about AB$_{1450}$ and the
EUV spectrum, so we expect that the true distributions of quasar
properties may only improve the suitability of the best SDSS quasar over
our estimates.

\begin{figure}
  \includegraphics[width=8.4cm]{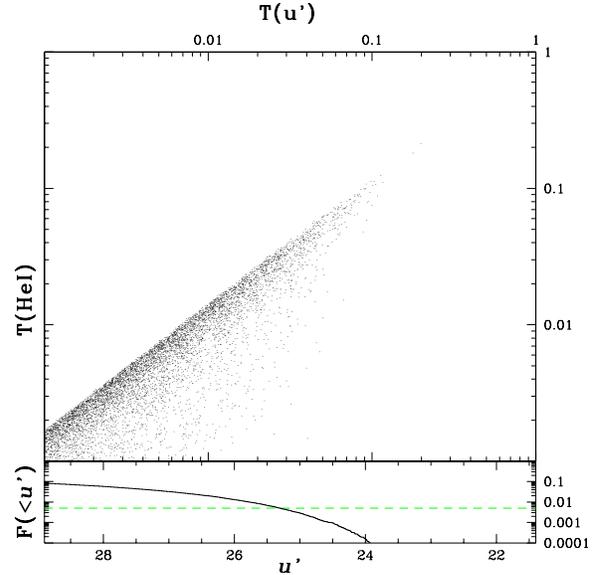}
  \caption{Plot of $T(\hei)$ versus $T(u')$.  This plot shows the mean
    transmissions in the \hei\ band and in the $u'$ band for each
    $z=5$ quasar LOS in $10^5$ LOS Monte Carlo simulations.  The
    excellent correlation between $T(\hei)$ and $T(u')$ at the high
    $T(\hei)$ end of the distribution shows that $u'$-band photometry
    is an effective method for selecting the best quasar for follow-up
    \hei-band spectroscopy.  Under the assumption that the intrinsic
    quasar spectrum is given by the model of Section
    \ref{sec:qsospec}, we convert $T(u')$ into a $u'$-band AB
    magnitude, and plot the cumulative fraction of quasars brighter
    than that magnitude in the bottom panel.  The horizontal dashed
    line at 1/200 intersects the curve at $u'=25.3$, the expected $u'$
    magnitude of the brightest quasar in a 200 quasar sample.}
  \label{fig:n2select4ub}
\end{figure}

For $z_\rmn{Q}=5$, the \hei\ band covers 3222 to 3506~\AA; this is
within the SDSS $u'$ filter, which runs from $\sim 3250$ to 3750~\AA\
\citep{fan01a}.  As a consequence we expect the $u'$ magnitude to
serve as a good proxy for the \hei-band flux.  Figure
\ref{fig:n2select4ub} shows the relationship between $T(\hei)$ and
$T(u')$, the transmitted flux fraction in the $u'$ band.  The strong
correlation illustrated in the scatter plot is independent of the
intrinsic quasar properties.  Assuming our standard quasar model, the
transmitted fractions are directly related to fluxes; the cumulative
histogram at the bottom of Fig.  \ref{fig:n2select4ub} shows the
fraction of observed quasars brighter than a given value of $u'$.  We
expect that 95 per cent of SDSS $z\simeq5$ quasars will be fainter
than $u'=27$, but the quasar with the highest \hei-band flux will have
$u'\sim25.5$, with an optimistic value of $u'=24$.  The $u'$-band
1-$\sigma$ limiting magnitude for the SDSS survey is expected to be
about $u'=24$ \citep{fan01a}, thus under optimistic assumptions there
is some chance the SDSS survey itself would make a $u'$ detection of a
$z\simeq5$ quasar, and thus identify an excellent candidate for
spectroscopic follow-up observations.  It is likely, however, that
deeper $u'$-band photometry on the SDSS quasars will be desirable to
locate the few brightest quasars in the $u'$ band.  The 2.5-meter SDSS
telescope integrates for only 54.1 seconds per filter, so achieving
deeper $u'$ photometry on a small telescope equipped with a
blue-sensitive CCD should not be difficult (extending the 1-$\sigma$
limit to $u'=25.5$ would only require 2.5 minutes per source with the
SDSS telescope, or about 8 hours of total integration time for all 200
SDSS $z\simeq5$ quasars).

Once candidates have been identified from $u'$-band photometry,
low-resolution spectroscopy should be performed to determine the flux
in the \hei\ band; because the $u'$ band extends longward of the \hei\
band, a quasar can be relatively bright in $u'$ but faint in the \hei\
band (note the points scattered below the main correlation in
Fig. \ref{fig:n2select4ub}).  The low-resolution spectroscopy will
require a blue-sensitive spectrograph, but on only a 2-meter class
telescope.

\begin{figure}
  \includegraphics[width=8.4cm]{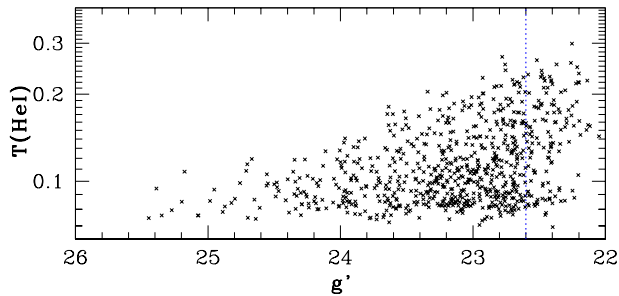}
  \caption{Plot of $T(\hei)$ versus $g'$ of quasars.  These points are
    for the standard intrinsic quasar flux model of Section
    \ref{sec:qsospec}.  Only quasars with $g'>22.60$ (plotted to the
    left of the dotted line) are identified by the SDSS quasar color
    selection algorithm as quasars.  Note that this plot focuses on
    quasars with very optimistic values of $T(\hei)\ga0.1$; quasars
    with lower values of $T(\hei)$ generally all fall at $g'>22.60$.}
  \label{fig:gcol}
\end{figure}

We now turn to our final point: the SDSS quasars are selected based on
photometry in all of the SDSS bands.  The primary SDSS
photometric-selection criteria for $z\simeq5$ quasars require
$u'>22.00$ and $g'>22.60$ \citep{fan01a}.  As illustrated in Fig.
\ref{fig:n2select4ub}, we do not expect any $z\simeq5$ quasars to be
nearly as bright as $u'=22.00$.  In contrast, Fig. \ref{fig:gcol}
shows that quasars with $T(\hei)=0.1$ will sometimes be slightly
brighter than $22.60$ in $g'$.  Though unlikely, it is possible that
the SDSS quasar selection algorithm could miss not only a $z\simeq5$
quasar, but the best one in the survey for our purposes.  When the
SDSS quasar survey is complete, and the SDSS data are publicly
released, a full search of the SDSS catalog can be made to determine
whether any objects meet all the SDSS $z\simeq5$ quasar criteria
except for $g'>22.60$, but still have, e.g., $g'>22.40$.  These
objects could be followed up independently.

\section{Summary}
\label{sec:summ}

We have developed a method to infer the dominant stellar population
responsible for the ionizing background radiation at $z\simeq5$, using
absorption lines in the spectrum of a $z\simeq5$ quasar.
Specifically, we related the strength of relative absorption in \hei\
lines compared to \hi\ lines to whether \popii\ or \popiii\ stars
dominated the ionizing background shortly after reionization
completed.

First we related the ionizing background shape to the ratio, $\eta$,
of \hei\ to \hi\ in an optically thin absorbing system; the spectrum
of \popii\ stars gives rise to $\eta=0.20$ while the spectrum of very
massive \popiii\ stars yields $\eta=0.06$.  Then we adopted a model
for the population of IGM absorbers based on absorption line
observations of quasars.  The existing observations do not tightly
constrain absorber populations at $z\ga4$; our model does extrapolate
to a neutral fraction of $1.4\times10^{-2}$ at $z=6.28$, consistent
with measurements.  We adopted a model quasar spectrum to shine
through our model IGM: we assumed the quasar flux to be at the faint
limit for the SDSS quasar survey (20th magnitude in the AB system at
rest-frame 1450~\AA), and modelled the EUV spectral shape based on the
composite spectrum of lower-redshift quasars with a spectral slope of
$\alpha_\mathrm{EUV}=1.6$.

We extrapolated the early results of the SDSS quasar survey to find
that SDSS will discover 200 quasars with $4.8<z<5.2$.  We used Monte
Carlo simulations of our IGM model to generate random realizations of
lines-of-sight (LOSs) toward $z_\rmn{Q}\simeq5$ quasars.  Based on
those results, we concluded that a sample of 200 quasars should
contain one quasar along a line-of-sight with 3 per cent transmission
by the IGM in the \hei\ band ($537(1+z)$ to $584(1+z)$~\AA); given the
uncertainty in the absorber model, we consider a LOS with 10 per cent
transmission to be an optimistic possibility.

We simulated large numbers of quasar LOSs with 3 and 10 per cent
transmission.  We placed a model quasar at $z=5$, and then simulated
observations of the \hei\ and \hi\ forests.  Based on the result for a
large ensemble of observations, we derived an estimation technique
that can discriminate (at 95 per cent confidence) a \popii\ ionizing
background from a \popiii\ ionizing background $\sim 50$ per cent of
the time for observations along our {\it expected} best LOS, and that
can discriminate the backgrounds 95 per cent of the time for
observations along our {\it optimistic} best LOS.  These observations
require very long integrations with a 10-meter class telescope
instrumented with a blue-sensitive low-dispersion spectrograph and an
optical intermediate-resolution spectrograph.

Finally, we examined the selection of the best quasar/LOS combination
on which to perform our proposed experiment from the expected 200 SDSS
quasars at $z\simeq5$.  The \hei\ forest at $z=5$ falls in the $u'$
filter; consequently $u'$ imaging is an efficient way to select the
best quasar/LOS combination.  However, the SDSS $u'$ photometry will
not be deep enough to detect $z\simeq5$ quasars (any detection would
be beyond the optimistic expectation of our IGM model), so follow-up
$u'$ photometry of the 200 $z\simeq5$ quasars will be required; these
observations will be inexpensive.  We also consider the possibility
that a quasar/LOS combination will be so good for our purposes that it
falls outside of the SDSS quasar color selection criteria.  This is
very unlikely for quasars with $\sim 10$ per cent or lower
transmission, but about half of all quasars with 20 per cent
transmission, should any exist, would be brighter in $g'$ than the
SDSS quasar color selection limit of 22.6.  If quasars are discovered
with 10 per cent transmission, it may be worthwhile to modify the SDSS
quasar color selection criteria so as to discover quasars with even
less IGM absorption in the \hei\ forest.

\section*{Acknowledgments}

MRS thanks the TA group at Harvard CfA for warm hospitality.  We
gratefully acknowledge helpful conversations with Wal Sargent, Rob
Simcoe, Chuck Steidel, Peng Oh and Marc Kamionkowski.  This work was
supported in part by NASA GSRP grant NGT5-50339 and NASA grant
NAG5-9821 (for MRS), and by NASA grant ATP02-0004-0093 and NSF grants
AST-0071019 and AST-0204514 (for AL).  AL acknowledges support from
the Institute for Advanced Study and the John Simon Guggenheim
Memorial Fellowship.

\bsp

\label{lastpage}


\begin{thebibliography}{99}
\bibitem[\protect\citeauthoryear{Anderson et al.}{2001}]{and01}
  Anderson S.F., et al., 2001, AJ, 122, 503
\bibitem[\protect\citeauthoryear{Barger et al.}{2003}]{barg03} Barger
  A.J., Cowie L.L., Capak P., Alexander D.M., Bauer F.E., Brandt W.N.,
  Garmire G.P., Hornschemeier A.E., 2003, ApJ, 584, L61
\bibitem[\protect\citeauthoryear{Barkana \& Loeb}{2001}]{bar01}
  Barkana R., Loeb A., 2001, Physics Reports, 349, 125
\bibitem[\protect\citeauthoryear{Bechtold}{1994}]{bec94}
  Bechtold J., 1994, ApJS, 91, 1
\bibitem[\protect\citeauthoryear{Bechtold}{2003}]{bec03}
  Bechtold J., 2003, in P\'erez-Fournon I., Balcells M., 
  Moreno-Insertis F., S\'anchez F., eds, Galaxies at High
  Redshift. Cambridge Univ. Press, Cambridge, p. 131 (astro-ph/0112521)
\bibitem[\protect\citeauthoryear{Becker et al.}{2001}]{bec01} 
  Becker R.H., et al., 2001, AJ, 122, 2850
\bibitem[\protect\citeauthoryear{Bromm, Kudritzki \& Loeb}{Bromm et al.}{2001}]{bro01}
  Bromm V., Kudritzki R.P., Loeb A., 2001, ApJ, 552, 464
\bibitem[\protect\citeauthoryear{Burles, Nollet \& Turner}{Burles et al.}{2001}]{bur01}
  Burles S., Nollet K.M., Turner M.S., 2001, ApJ, 552, L1
\bibitem[\protect\citeauthoryear{Cen}{2003a}]{cen03a}
  Cen R., 2003a, ApJ, submitted, astro-ph/0210473
\bibitem[\protect\citeauthoryear{Cen}{2003b}]{cen03b}
  Cen R., 2003b, ApJ, submitted, astro-ph/0303236
\bibitem[\protect\citeauthoryear{Djorgovski et al.}{2001}]{djo01} 
  Djorgovski S.G., Castro S., Stern D., Mahabal A.A., 2001, ApJ, 560, L5
\bibitem[\protect\citeauthoryear{Dobrzycki et al.}{2002}]{dob02}
  Dobrzycki A., Bechtold J., Scott J., Morita M., 2002, ApJ, 571, 654
\bibitem[\protect\citeauthoryear{Fan et al.}{1999}]{fan99} 
  Fan X., et al., 1999, AJ, 118, 1
\bibitem[\protect\citeauthoryear{Fan et al.}{2001a}]{fan01a}
  Fan X., et al., 2001a, AJ, 121, 31
\bibitem[\protect\citeauthoryear{Fan et al.}{2001b}]{fan01b} 
  Fan X., et al., 2001b, AJ, 121, 54
\bibitem[\protect\citeauthoryear{Fan et al.}{2002}]{fan02} 
  Fan X., et al., 2002, AJ, 123, 1247
\bibitem[\protect\citeauthoryear{Haardt \& Madau}{1996}]{haa96} 
  Haardt F., Madau P., 1996, ApJ, 461, 20
\bibitem[\protect\citeauthoryear{Hewitt \& Burbidge}{1987}]{hew87}
  Hewitt A., Burbidge G., 1987, ApJS, 63, 1
\bibitem[\protect\citeauthoryear{Jakobsen et al.}{1994}]{jak94}
  Jakobsen P., Boksenberg A., Deharveng J.M., Greenfield P.,
  Jedrzejewski R., Paresce F., 1994, Nature, 370, 35
\bibitem[\protect\citeauthoryear{Jakobsen}{1998}]{jak98}
  Jakobsen P., 1998, A\&A, 335, 876
\bibitem[\protect\citeauthoryear{Kim et al.}{1997}]{kim97}
  Kim T.-S., Hu E.M., Cowie L.L., Songaila A., 1997, AJ, 114, 1
\bibitem[\protect\citeauthoryear{Kriss et al.}{2001}]{kri01}
  Kriss G.A., et al., 2001, Science, 293, 1112
\bibitem[\protect\citeauthoryear{Kogut et al.}{2003}]{kog03}
  Kogut A., et al., 2003, ApJ, submitted, astro-ph/0302213
\bibitem[\protect\citeauthoryear{Leitherer et al.}{1999}]{lei99}
  Leitherer C., et al., 1999, ApJS, 123, 3
\bibitem[\protect\citeauthoryear{Liske et al.}{2001}]{lis00}
  Liske J., Webb J.K., Williger G.M., Fern\'{a}ndez-Soto A., Carswell R.F., 
  2000, MNRAS, 311, 657
\bibitem[\protect\citeauthoryear{Loeb \& Barkana}{2001}]{loe01}
  Loeb A., Barkana R., 2001, ARA\&A, 39, 19
\bibitem[\protect\citeauthoryear{McCarthy et al.}{1998}]{mcc98}
  McCarthy J.K., et al., 1998, in D'odorico S., ed, Proc. SPIE
  Vol. 3355, Optical Astronomical Instrumentation. p. 81
\bibitem[\protect\citeauthoryear{Miralda-Escud\'e \& Ostriker}{1992}]{mir92}
  Miralda-Escud\'e J., Ostriker J.P., 1992, ApJ, 392, 15
\bibitem[\protect\citeauthoryear{M\o ller \& Jakobsen}{1990}]{mol90}
  M\o ller P., Jakobsen P., 1990, A\&A, 228, 299
\bibitem[\protect\citeauthoryear{Oh}{2001}]{oh01}
  Oh S.P., 2001, ApJ, 553, 499
\bibitem[\protect\citeauthoryear{Oke et al.}{1995}]{oke95}
  Oke J.B., et al., 1995, PASP, 107, 375
\bibitem[\protect\citeauthoryear{Osterbrock}{1989}]{ost89}
  Osterbrock D.E., 1989, Astrophysics of Gaseous Nebulae and Active
  Galactic Nuclei. University Science Books, Sausalito, CA
\bibitem[\protect\citeauthoryear{Peebles}{1993}]{pee93}
  Peebles P.J.E., 1993, Principles of Physical Cosmology. Princeton
  University Press, Princeton, NJ
\bibitem[\protect\citeauthoryear{Prochaska et al.}{2001}]{pro01}
  Prochaska, J.X., et al., 2001, ApJS, 137, 21
\bibitem[\protect\citeauthoryear{Sargent, Steidel, \& Boksenberg}{Sargent et al.}{1989}]{sar89}
  Sargent W.L.W., Steidel C.C., Boksenberg A., 1989, ApJS, 69, 703
\bibitem[\protect\citeauthoryear{Scheffler \& Els\"asser}{1988}]{sch88}
  Scheffler H., Els\"asser H., 1988, Physics of the Galaxy and Interstellar Matter. Springer-Verlag, New York
\bibitem[\protect\citeauthoryear{Schneider et al.}{2002}]{sch02}
  Schneider D.P., et al., 2002, AJ, 123, 567
\bibitem[\protect\citeauthoryear{Scott et al.}{2000}]{sco00}
  Scott J., Bechtold J., Morita M., Dobrzycki A., Kulkarni V.P., 2002, ApJ, 571, 665
\bibitem[\protect\citeauthoryear{Sheinis et al.}{2002}]{she02}
  Sheinis A.I., Bolte M., Epps H.W., Kibrick R.I., Miller J.S.,
  Radovan M.V., Bigelow B.C., Sutin B.M., 2002, PASP, 114, 851
\bibitem[\protect\citeauthoryear{Spergel et al.}{2003}]{spe03}
  Spergel D.N., et al., 2003, ApJ, submitted, astro-ph/0302209
\bibitem[\protect\citeauthoryear{Stengler-Larrea et al.}{1995}]{ste95}
  Stengler-Larrea E.A., et al., 1995, ApJ, 444, 64
\bibitem[\protect\citeauthoryear{Storrie-Lombardi et al.}{1994}]{sto94}
  Storrie-Lombardi L.J., McMahon R.G., Irwin M.J., Hazard C., 1994, ApJ, 427, L13
\bibitem[\protect\citeauthoryear{Storrie-Lombardi \& Wolfe}{2000}]{sto00}
  Storrie-Lombardi L.J., Wolfe A.M., 2000, ApJ, 543, 552
\bibitem[\protect\citeauthoryear{Telfer et al.}{2002}]{tel02}
  Telfer R.C., Zheng W., Kriss G.A., Davidsen A.F., 2002, ApJ, 565, 773
\bibitem[\protect\citeauthoryear{Theuns et al.}{2002}]{the02} 
  Theuns T., Bernardi M., Frieman J., Hewett P., Schaye J., Sheth R.~K., Subbarao M., 2002, ApJ, 574, L111
\bibitem[\protect\citeauthoryear{Vanden Berk et al.}{2001}]{van01}
  Vanden Berk D.E., et al., 2001, AJ, 122, 549
\bibitem[\protect\citeauthoryear{Verner et al.}{1996}]{ver96}
  Verner D.A., Ferland G.J., Korista K.T., Yakovlev D.G., 1996, ApJ, 465, 487
\bibitem[\protect\citeauthoryear{Weymann et al.}{1998}]{wey98}
  Weymann R., et al., 1998, ApJ, 506, 1
\bibitem[\protect\citeauthoryear{White et al.}{2003}]{whi03}
  White R.L., Becker R.H., Fan X., Strauss M.A., 2003, AJ, submitted, astro-ph/0303476
\bibitem[\protect\citeauthoryear{Wyithe \& Loeb}{2003a}]{wyi03a}
  Wyithe J.S.B., Loeb, A., 2003, ApJ, 586, 693
\bibitem[\protect\citeauthoryear{Wyithe \& Loeb}{2003b}]{wyi03b}
  Wyithe J.S.B., Loeb, A., 2003, ApJ, submitted, astro-ph/0302297
\bibitem[\protect\citeauthoryear{Zheng et al.}{1997}]{zhe97}
  Zheng W., Kriss G.A., Telfer R.C., Grimes J.P., Davidsen A.F., 1997, ApJ, 475, 469
\bibitem[\protect\citeauthoryear{Zuo \& Phinney}{1993}]{zuo93}
  Zuo L., Phinney E.S., 1993, ApJ, 418, 28

\end{thebibliography}
\end{document}